\documentclass[prl, reprint, amsmath, amssymb, floatfix, aps, longbibliography, superscriptaddress]{revtex4-2}
\usepackage{graphicx}
\usepackage{dcolumn}
\usepackage{bm}
\usepackage[colorlinks=true,urlcolor=rossos,anchorcolor=black,citecolor=mygreen,filecolor=black,linkcolor=black,menucolor=blue,linktocpage=true]{hyperref} 
\usepackage{subfigure}

\usepackage{booktabs}
\usepackage{colortbl}
\usepackage{collcell}
\usepackage{enumerate}
\usepackage{float}
\usepackage{verbatim}
\usepackage{multirow}
\usepackage{xparse}
\usepackage{tabularx}
\usepackage{cancel}

\usepackage{multirow}

\usepackage[normalem]{ulem}
\usepackage{soul}
\setstcolor{red}

\definecolor{rossos}{cmyk}{0,1,1,0.55}
\definecolor{mygreen}{rgb}{0.27, 0.64, 0.48}
\definecolor{mygray}{gray}{0.95}

\begin{document}

\title{High-Quality Axions in a Class of Chiral $U(1)$ Gauge Theories}

\author{Yu-Cheng Qiu}
\email{ethanqiu@sjtu.edu.cn}
\affiliation{Tsung-Dao Lee Institute and School of Physics and Astronomy, \\
Shanghai Jiao Tong University, 520 Shengrong Road, Shanghai, 201210, China}

\author{Jin-Wei Wang}
\email{jinwei.wang@uestc.edu.cn}
\affiliation{School of Physics, University of Electronic Science and Technology of China, Chengdu 611731, China}
\affiliation{Tsung-Dao Lee Institute and School of Physics and Astronomy, \\
Shanghai Jiao Tong University, 520 Shengrong Road, Shanghai, 201210, China}

\author{Tsutomu T. Yanagida}
\email{tsutomu.tyanagida@sjtu.edu.cn}
\affiliation{Tsung-Dao Lee Institute and School of Physics and Astronomy, \\
Shanghai Jiao Tong University, 520 Shengrong Road, Shanghai, 201210, China}
\affiliation{Kavli IPMU (WPI), The University of Tokyo, Kashiwa, Chiba 277-8583, Japan}

\begin{abstract}
We show that there are many candidates for the quintessence and/or the QCD axions in a class of chiral $U(1)$ gauge theories. Their qualities are high enough to serve as the dark energy and/or to solve the strong $CP$ problem. Interestingly, the high quality of axion is guaranteed by the gauged $U(1)$ and $\mathbf{Z}_{2N}$ symmetries and hence free from the nonperturbative quantum gravity corrections. Furthermore, our mechanism can be easily applied to the Fuzzy dark matter axion scenarios.
\end{abstract}

\maketitle

\noindent \textit{\textbf{Introduction.}}---The observed cosmological constant (CC), $\Lambda\simeq (2.26\times 10^{-3}\,{\rm eV})^4$~\cite{Planck:2018vyg}, is one of biggest mysteries in nature. One may ask a natural question: is it a constant or potential energy of a scalar boson field? 

In this Letter, we stick to the latter scenario, because if so, it may provide us a deep insight into the quantum gravity~\cite{Dvali:2018fqu, Obied:2018sgi,tHooft:2006uhw,Lin:2022khg}. In this case, the mass of the scalar boson must be assumed extremely small as $\sim 10^{-33}\,{\rm eV}$ in order to keep the boson at the nonminimum point of its potential until the present. A unique candidate is the Nambu-Goldstone boson (called here as a quintessence axion~\cite{Fukugita:1994hq,Fukugita:1995nb,Frieman:1995pm,Kolda:1998wq,Choi:1999wv,Kim:2002tq,Kaloper:2005aj,Bonnefoy:2018ibr}) since it can have such a small mass against possible radiative corrections. However, the nonperturbative corrections of the quantum gravity may easily generate a larger mass for the axion, since nonperturbative corrections explicitly break any global symmetry in the quantum gravity~\cite{Banks:2010zn}. If it happens, the axion is no longer able to explain the present CC. We call this problem the quality problem of quintessence axion.

Interestingly, there is another candidate for a light particle, that is, the QCD axion. The QCD axion~\cite{Wilczek:1977pj, Weinberg:1977ma} has attracted many people's attention for a long time since it provides us a dynamical solution to the strong $CP$ problem \cite{Peccei:1977hh}. However, due to the stringent constraint on QCD vacuum angle from neutron EDM measurement, the QCD axion also faces a similar quality problem~\cite{Georgi:1981pu,Barr:1992qq,Kamionkowski:1992mf,Holman:1992us,Ardu:2020qmo}.

Another issue is that the origin of both axions in UV theories is unknown. String theories are expected to be such UV theories, and in fact, there are many candidates for massless axions whose masslessness is guaranteed by shift symmetries at the tree level in string theories. 
However, world-sheet instantons and/or gravitational instantons might generate huge breakings of the shift symmetries~\cite{Svrcek:2006yi} and if it is the case the axions do not remain at low energies. 
Therefore, it is very important to search for the UV theories in the framework of quantum field theories~\cite{Fukuda:2017ylt,Ibe:2018hir,Choi:2022fha}.

In this Letter, we point out that candidates for the quintessence and QCD axions often exist in large parameter space for a class of the chiral $U(1)$ gauge theories. Surprisingly, the quality of the axions required to explain the observed vacuum energy (equivalently the CC) and/or to solve the strong $CP$ problem is guaranteed by the gauged $U(1)$ and $\mathbf{Z}_{2N}$ symmetries~\footnote{ 
The gauged discrete symmetries have been widely used to get rid of dangerous lower-order operators and achieve stable axion solutions~\cite{Dias:2002gg,Dias:2002hz,Chun:1992bn,Bastero-Gil:1997brz,Babu:2002ic}. By contrast, the more important function of gauge $\mathbf{Z}_{2N}$ symmetry in our Letter is to prohibit the unwanted fermion loops [see the content around Eq.~\eqref{eq:PQbreak} for more details].}. Moreover, our mechanism can also be extended to include the Fuzzy dark matter (DM) axion scenario.\\

\noindent \textit{\textbf{Chiral $U(1)$ gauge theories.}}---The new sector consists of two Higgs $\phi_1$, $\phi_2$ and $N$ pairs of chiral fermions $\{\psi_i,\overline{\psi}_i\}$, where $i=1,2,\cdots,N$. 
Two Higgs fields imply two global $U(1)$ symmetries associated with their phase rotations. As shown in Ref.~\cite{Fukuda:2017ylt}, one linear combination of two $U(1)$s can be gauged dubbed $U(1)_g$, while the other combination dubbed $U(1)_a$, is orthogonal to $U(1)_g$ and can be the origin of the axion. 

Since $U(1)_g$ is a gauge symmetry, there are two anomaly-cancellation conditions must be fulfilled, which are from $[U(1)_g]^3$ and gravitational $[U(1)_g]\times [{\rm graviton}]^2$ anomaly, i.e.,
\begin{subequations}
\label{eq:anomaly_cancel}
\begin{align}
    \sum_{i=1}^N U(1)_g^{\psi_i} + U(1)_g^{\overline{\psi}_i} & = 0\;,  \label{eq:anomali_cancel_1} \\
    \sum_{i=1}^N \left[U(1)_g^{\psi_i}\right]^3 + \left[U(1)_g^{\overline{\psi}_i}\right]^3 & = 0\;,
    \label{eq:anomaly_cancel_2}
\end{align}
\end{subequations}
where $U(1)_g^{\psi_i}$ ($U(1)_g^{\overline{\psi}_i}$) represents the $U(1)_g$ charge of $\psi_i$ ($\overline{\psi}_i$).  Note that all these charges should be rational numbers, otherwise, it violates a principle in the quantum gravity~\cite{Banks:2010zn}. Furthermore, we can make them all integers by proper normalization. In addition, the assignment of $U(1)_g^{\psi_i}$ and $U(1)_g^{\overline{\psi}_i}$ needs to ensure that there is no gauge invariant mass term, otherwise, they get the Planck-scale masses and become irrelevant at low energies. 
We demand that all fermions acquire mass only through the Yukawa couplings. Therefore, the $U(1)_g$ charge of two Higgs $q_{1,2}$ can be determined by gauge invariance.   

\begin{table*}[t]
\caption{Fermion charge assignment.}
\label{tab:asymmetriccharges}
\begin{ruledtabular}
\begin{tabular}{ccccccccccccccccc}
$i$ & $1$ & $2$ & $3$ & $4$ & $\cdots$ &  & & & $k+1$ & $k+2$ & $\cdots$  & & & &  &   \\
\midrule
$\psi_i$ & $\alpha_1$ & $\beta_1$ &  $\alpha_2$ & $\beta_2$ & $\cdots$ & $\gamma_1$ & $\gamma_2$ & $\cdots$ & $\delta_1$ & $\eta_1$ & $\delta_2$ & $\eta_2$ & $\cdots$  & $\sigma_1$ & $\sigma_2$  & $\cdots$ \\[0.4em]
$\overline{\psi}_i$ & $\beta_1$ & $\alpha_1$ &  $\beta_2$ & $\alpha_2$ & $\cdots$ & $\gamma_1$ & $\gamma_2$ & $\cdots$ & $\eta_1$ & $\delta_1$ & $\eta_2$ & $\delta_2$ & $\cdots$  & $\sigma_1$ & $\sigma_2$  & $\cdots$ \\
\end{tabular}
\end{ruledtabular}
\end{table*}

Assume that $k$ pairs of fermions couple to $\phi_1$ and the rest $l=N-k$ pairs of fermions couple to $\phi_2$~\footnote{In general, there are no known analytical methods to solve Eq.~\eqref{eq:anomaly_cancel}--\eqref{eq:higgs_charge} for an arbitrary value of $k$ and $l$~\cite{Batra:2005rh}. However, in the following content, we will show that we can use some tricks to find suitable solutions.}. The corresponding $U(1)_g$ charge of these fermions are shown in Table~\ref{tab:asymmetriccharges}. Then we have
\begin{subequations}
\begin{align}
    -q_1 & = \alpha_1+\beta_1 = \alpha_2+\beta_2=\cdots = 2\gamma_1=\cdots\;,\\
    -q_2 & = \delta_1+ \eta_1 = \delta_2 + \eta_2 =\cdots = 2\sigma_1 =\cdots \;.
\end{align}\label{eq:higgs_charge}
\end{subequations}
Note that the  $\psi_i$ and $\overline{\psi}_i$ carry the same $U(1)_g$ charge but in a different order, which can reduce the number of degrees of freedom and make it much easier to solve Eq. \eqref{eq:anomali_cancel_1} and \eqref{eq:anomaly_cancel_2}. According to Eq.~\eqref{eq:anomali_cancel_1}, one can obtain that
\begin{equation}
    -\frac{q_1}{q_2} = \frac{l}{k} = \frac{m}{n}\;,
\end{equation}
where $m$ and $n$ are relatively prime integers.
Without loss of generality, we set $q_1>0$ and $q_2<0$. With $N$ pairs of new chiral fermions and two Higgs bosons $\phi_{1,2}$, we find that there is an interesting accidental discrete symmetry, that is $\mathbf{Z}_{2N}$, under which $\psi_i$ and $\overline{\psi}_i$ both carry charge $1$ and $\phi_{1,2}$ carries charge $-2$. Besides, it is straightforward to check that this $\mathbf{Z}_{2N}$ is anomaly-free, and therefore we could regard it as a gauge discrete symmetry \cite{Ibanez:1991hv}. We will see below that this gauged $\mathbf{Z}_{2N}$ is crucial to our results.

As we mentioned above, high quality is extremely crucial for both quintessence and QCD axion, that is, the global $U(1)_a$ should be a good symmetry. 
In our framework, the possible lowest-order nonrenormalizable operator that obeys the gauge $U(1)_g$ and $\mathbf{Z}_{2N}$ symmetry but breaks the global $U(1)_a$ symmetry is
\begin{equation}
\mathcal{O} = \frac{1}{k!l!}\frac{\phi_1^k \phi_2^l}{M_{\rm Pl}^{N-4}} + \text{h.c.}\;,
\label{eq:PQbreak}
\end{equation}
where $M_{\rm Pl} = 2.4 \times 10^{18}\,{\rm GeV}$ is the reduced Planck scale.
Clearly, varying degrees of qualities can be achieved by adjusting the values of $k$ and $l$. 

Note that the correctness of Eq.~\eqref{eq:PQbreak} is based on the pointlike interaction assumption, however, the potential fermion loops that are induced by higher-order operators, e.g. $\phi_{1,2}^r (\psi_i \overline{\psi}_j)^s$, may also render $\mathcal{O}$-like operators by integrating out the heavy fermions. In this case, the $\mathcal{O}$-like operators will receive a lower suppression, e.g. $\mathcal{O}\sim \phi_1^k \phi_2^l/m_{\psi_i}^{N-4}$. Thanks to the gauged $\mathbf{Z}_{2N}$ symmetry, the dimension of allowed higher-order operators will be much higher, e.g.  $\phi_{1,2}^r (\psi_i \overline{\psi}_j)^{N+r}$ and/or $\phi_{1,2}^{*r} (\psi_i \overline{\psi}_j)^{N-r}$. By doing a simple dimension analysis one can find that the contribution of the fermion loop is highly suppressed compared with Eq.~\eqref{eq:PQbreak}. Therefore, in the following content, we will stick to Eq.~\eqref{eq:PQbreak} \footnote{Note that for a specific model, i.e. with a fixed number of fermions and $U(1)_g$ charge assignment, these dangerous fermion loops may not exist even without gauged $\mathbf{Z}_{2N}$, but it is difficult to give a rigorous mathematical proof. However, by adopting gauged $\mathbf{Z}_{2N}$, we can guarantee the correctness of Eq.~(4). Therefore, in our framework, we use both gauge $U(1)_g$ as well as $\mathbf{Z}_{2N}$ symmetries to make sure our results are more robust.}.

After spontaneous symmetry breaking, one could expand two Higgs fields as $\phi_1 = (f_1/\sqrt{2}) \exp{(i\tilde{a}/f_1)}$ and $\phi_2 = (f_2/\sqrt{2}) \exp{(i \tilde{b}/f_2)}$, where $f_i$ is the vacuum expectation value of $\phi_i$. Since here we focus on two Nambu-Goldstone modes $\tilde{a}$ and $\tilde{b}$, the radial modes are neglected. One linear combination of them, $b$, is absorbed by the gauge boson of $U(1)_g$, while the orthogonal mode, $a$, is the axion. They are related by~\cite{Fukuda:2017ylt}
\begin{equation}
\begin{pmatrix}
a\\
b
\end{pmatrix}
= \frac{-1}{\sqrt{q_1^2 f_1^2 + q_2^2 f_2^2}}
\begin{pmatrix}
q_2f_2 & -q_1 f_1 \\ 
q_1f_1 & q_2 f_2
\end{pmatrix}
\begin{pmatrix}
\tilde{a}\\
\tilde{b}
\end{pmatrix}\;.
\label{eq:transform}
\end{equation}
Therefore, one has
\begin{equation}
\phi_1^k \phi_2^{l} = \frac{f_1^k f_2^l}{2^{N/2}} e^{(i a N_{\rm DW}  / F_a)}\;, \quad F_a = \frac{f_1 f_2}{\sqrt{m^2 f_1^2 + n^2 f_2^2}}\;,
\label{eq:Fa}
\end{equation}
where $N_{\rm DW}$ is the domain wall number, which happens to be the greatest common divisor of $(k,l)$, so we have $(n,m)=(k/N_{\rm DW},l/N_{\rm DW})$ (see Supplemental Material for details).
Clearly, $\mathcal{O}$ breaks the continuous shift symmetry of $a$ and contributes to the axion mass. The $b$ mode does not show up in $\mathcal{O}$ as expected since it is $U(1)_g$ invariant.
In the following content, we will show that this formalism can always provide us with a proper quintessence axion and/or QCD axion candidate.\\


\noindent \textit{\textbf{High-quality quintessence axion.}}---
Now we construct the quintessence axion to explain the observed CC.
Assuming that $\psi_i\in(1,\mathbf{2},0)$ and $\overline{\psi}_i\in(1,\mathbf{2^*},0)$ under the $SU(3)_{\rm c}\times SU(2)_{\rm L} \times U(1)_{\rm Y}$ gauge transformation. Then we can prove that axion has the Chern-Simons coupling (as shown in the Supplemental Material), that is
\begin{equation}
    \mathcal{L} \supset  N_{\rm DW}\frac{a}{F_a}\frac{g_2^2} {32\pi^2}W_{\mu\nu}^a\tilde{W}^{\mu\nu a}\;,
    \label{eq:CS_coupling}
\end{equation}
where $W_{\mu\nu}^a$ and $\tilde{W}^{\mu\nu a}$ are $SU(2)_L$ gauge field strength and its dual. In principle, both the high-order operator $\mathcal{O}$ and the $SU(2)_L$ instanton effect can contribute to the axion potential. However, as shown in Ref.~\cite{Nomura:2000yk}, without supersymmetry the contribution of $SU(2)_L$ instanton is negligible. 
Thus, the axion potential is only generated from the higher-order symmetry-breaking operator (see Eq.~\eqref{eq:PQbreak}), which can be expressed as
\begin{equation}
V  = \frac{\Lambda_a}{2} \left( 1- \cos{\frac{aN_{\rm DW}}{F_a}}\right)\;,
\label{eq:V}
\end{equation}
where
\begin{equation}
    \Lambda_a = \frac{2^{2-N/2}}{k! l!} \frac{f_1^k f_2^l}{M_{\rm Pl}^{N-4}}
    \label{eq:lambdaaxion}
\end{equation}
represents the potential energy at the hilltop, and can be used to explain the observed CC.

The equation of motion of the axion within the Friedmann–Lemaître–Robertson–Walker metric is given by
\begin{equation}
\ddot{a} + 3H(t) \dot{a} + \partial_a V =0\;,
\end{equation}
where $H(t)$ is the Hubble constant and the dot refers to the derivative with respect to cosmic time, $t$. The curvature of the axion potential and Hubble constant determines the evolution of the axion. 
The quintessence axion requires the $\partial_a V$ small enough so that it is still frozen by the current Hubble constant, $H_0\sim 10^{-33}\,{\rm eV}$ or just starts to roll down toward its vacuum. 
If the initial field value of axion, is around the minimum, the curvature is determined by the axion mass, $\partial_a V \sim m_a^2 a$, which is too large to fulfill the slow-roll condition, 
\begin{equation}
m_a \sim \sqrt{\frac{\Lambda_a N_{\rm DW}}{F_a^2}} > \sqrt{N_{\rm DW}} \times 10^{-33}\,{\rm eV}\;,
\end{equation}
for $F_a\lesssim M_{\rm Pl}$ and $\Lambda_a=\Lambda$. Therefore, we need to put the axion around the hilltop initially. This brings us to the instability problem, which requires that one has a large enough $F_a$~\cite{Ibe:2018ffn,Choi:2021aze}.

We consider that the quintessence axion has good quality if 
\begin{equation}
\frac{F_a}{N_{\rm DW}} > 10^{16} \,{\rm GeV}\;, \quad 10^{-2} \Lambda \lesssim \Lambda_a \lesssim 10^{2} \Lambda\;.
\label{eq:ma_range}
\end{equation}
To quantitatively discuss the quality of quintessence axion, here we take $f_2 = f_1 = M_{\rm Pl}$ as a benchmark,  
which gives $\Lambda_a = 4M_{\rm Pl}^4/(2^{N/2} k!l!)$. In this case, the presence of $1/k!l!$ ensures the validity of the expansion of Eq.~\eqref{eq:PQbreak}.
Apparently, to explain the observed CC, one needs large $k$ and $l$.
Note that there are several ways to better solve the instability problem, for example, by setting a higher cutoff in Eq.~\eqref{eq:PQbreak}, say Planck scale $1.2 \times 10^{19}\,{\rm GeV}$, we can have $F_a\sim 10^{17}\,{\rm GeV}$, and this will make our quintessence axion scenario more robust.
Besides, one could also couple this axion with an extra gauge field so that the axion could achieve the slow roll with additional friction besides the Hubble, as adopted in natural inflation~\cite{Freese:1990rb,Adshead:2012kp}.

Solving Eq.\eqref{eq:lambdaaxion} and \eqref{eq:ma_range}, we can find many combinations of $k$ and $l$ that can provide appropriate quintessence axion, for example when $k=4$ and $l=74$, we have $\Lambda_a=1.17\Lambda$ and $F_a/N_\text{DW} = 3.24\times 10^{16}\,{\rm GeV}$. However, with such large $k$ and $l$, it is extremely difficult to solve the Eq.\eqref{eq:anomaly_cancel}--\eqref{eq:higgs_charge}. 
Here we use a trick to overcome this problem. First assume $k'$ and $l'$ pairs of fermions, where $k'$ and $l'$ are relatively small and mutually prime numbers. It is much easier to derive the fermion charge assignment by solving Eq.~\eqref{eq:anomaly_cancel}--\eqref{eq:higgs_charge}. Then, do $p$ copies (similar to the concept of generation in the Standard Model) to get the final $k=p k'$ and $l=p l'$ pairs of fermions as long as Eq.~\eqref{eq:ma_range} can be fulfilled. 
In fact, we could identify $N_{\rm DW} = p$ if there are appropriate solutions.
Here we give one specific example, taking $k'=1$, $l'=9$, and $p=8$, and the corresponding $U(1)_g$ charges are \{$-27, 5, 1, 15, -9, 19, -13, 29, -23, 3$\}. Then we can derive that $k=8$, $l=72$, $F_a/N_\text{DW}=3.3\times10^{16}\,{\rm GeV}$, and $\Lambda_a=1.88\Lambda$. Because of the large value of $F_a$ (above inflation scale) there is no domain wall problem even with $N_\text{DW}=8$. Furthermore, although there are many pairs of new fermions, we have checked that the $g_2$ will not reach Landau pole because of the large value of $f_1$ and $f_2$. This conclusion also holds true in the case of Fuzzy DM axion discussed below.

The presence of CS coupling~\eqref{eq:CS_coupling} indicates the coupling between the quintessence axion and photon after EW symmetry breaking,
\begin{equation}
  \mathcal{L} \supset N_{\rm DW} \frac{a}{F_a} \frac{g^2 }{32\pi^2} F_{\mu\nu}\tilde{F}^{\mu\nu}\;,
  \label{eq:photon_coupling}
\end{equation}
where $F_{\mu\nu}$ and $\tilde{F}^{\mu\nu}$ are photon field strength and its dual. Here $g$ has absorbed the electroweak mixing angle. As shown in Ref.~\cite{Choi:2021aze,Lin:2022niw}, this quintessence axion could be used to explain the isotropic cosmic birefringence.\\


\noindent \textit{\textbf{High-quality Fuzzy dark matter axion.}}--- The Fuzzy DM of mass $10^{-21}$--$10^{-19}\,{\rm eV}$~\cite{Irsic:2017yje,Armengaud:2017nkf,Ferreira:2020fam,Hui:2021tkt} is very attractive, since we may naively understand the size of galaxies by its de Broglie wavelength. Furthermore, it may not have small-scale problems including the cusp-core problem. Interestingly, the required initial value of the Fuzzy DM field to explain the DM density by its coherent oscillation is about $F_a\simeq 10^{16}\,{\rm GeV}$ which is close to the decay constant for the quintessence axion discussed above~\footnote{A recent proposal of mixed Fuzzy and cold DM model is constructed from electroweak axions~\cite{Qiu:2022uvt}.}. Thus, it is natural to accommodate both axions together in the present framework. 
It is in fact possible if we introduce a new set of fermions and Higgs that coupled to a new chiral $U(1)_g'$ and $\mathbf{Z}_{2N}'$ gauge symmetry. Thus, operator mixing among Higgs fields is avoided.

Here for Fuzzy DM, good quality means that the axion has suitable mass, $10^{-21}$--$10^{-19}\,{\rm eV}$, and we take $F_a=10^{16}\,{\rm GeV}$ as the benchmark.
Expanding the axion potential around the minimum (see Eq.~\eqref{eq:V}), the axion mass can be expressed as  
\begin{equation}
m_a = N_{\rm DW} M_{\rm Pl} \sqrt{ \frac{2}{k!l!}} \left( \frac{m^2+n^2}{2}\right)^{N/4} \left(\frac{F_a}{M_{\rm Pl}}\right)^{N/2-1}\;,
\label{eq:ma}
\end{equation}
where we have taken $f_1=f_2$ for simplicity.
As expected, large $k$ and $l$ are required to have a light mass.
Using the same trick as the quintessence axion, one could find a set of fermion charges that gives rise to a good quality Fuzzy DM axion. For example, taking $k'=1$, $l'=6$, and $p=7$, and the corresponding $U(1)_g$ charges are \{$-21, 4, 3, 16, -9, 20, -13$\}. Then we can derive that $k=7$, $l=42$, $m_a=2.5\times10^{-20}\,{\rm eV}$.
The Fuzzy DM axion under this framework has the same CS-type interaction with photon as Eq.~\eqref{eq:photon_coupling}, which provides a channel for future detection.\\


\noindent \textit{\textbf{High-quality QCD axion.}}--- 
The QCD axion model was proposed based on a chiral $U(1)_g$ gauge theory, where five pairs ($N=5$) of chiral quarks, $Q_i$ and $\overline{Q}_i$, have ``asymmetric" $U(1)_g$ charges. A known example is $\{-9,-5,-1,7,8\}$ for both $Q_i$ and $\overline{Q}_i$, where all gauge anomalies are canceled out~\cite{Nakayama:2011dj}. Two Higgs $\phi_{1,2}$ carry the $U(1)$ gauge charges $10$ and $-15$ to give masses to all fermions~\cite{Choi:2020vgb}. This is a consistent model for the QCD axion, since the axion couples to the QCD Chern-Simons term. However, the quality is not sufficiently high to solve the strong $CP$ problem~\footnote{An extremely high-quality QCD axion model was, recently, constructed based on this five-pair fermion model with help of supersymmetry and $R$ symmetries~\cite{Choi:2022fha}.}.

In this section, we extend the above model by introducing more fermions to get a high-quality QCD axion under this framework. There might be various extensions to solve the quality problem. 
Here we consider only a special case where we have $N=3+2x$ pairs of chiral fermions, $\psi''_i \in (\mathbf{3},1,0)$ and $\overline{\psi''}_i\in (\mathbf{3^*},1,0)$ under $SU(3)_{\rm c} \times SU(2)_{\rm L} \times U(1)_{\rm Y}$. Their $U(1)_g$ charge assignment is the special case in Table~\ref{tab:asymmetriccharges} with $k=3$ and $l=2x$, where there is only one $\gamma$ and no $\sigma_i$. 
The $U(1)_g$ charges of two Higgs $q_1$ and $q_2$ now obey $-q_1/q_2 = 2x/3$. Here we assume that $2x$ and $3$ are relatively prime numbers, and there is no domain wall problem (see Supplemental Material for details).

The high-order operator in Eq.~\eqref{eq:PQbreak} will cause a shift of the global minimum of axion potential, and therefore contribute to the QCD $\bar{\theta}$, i.e.,
\begin{equation}
\label{eq:delttheta}
\delta\bar{\theta} \sim \frac{2}{3!(2x)!} \left( \frac{F_a}{M_{\rm Pl}}\right)^{N}  \left(\frac{9+4x^2}{2}\right)^{N/2} \frac{M_{\rm Pl}^4}{m_\pi^2 F_\pi^2}\;,
\end{equation}
where $m_\pi$ and $F_\pi$ are the mass and decay constant of the pion. 
Here $f_1=f_2$ is also assumed.
In order to fulfill the high-quality requirements, we need $\delta\bar{\theta}<10^{-10}$~\cite{Pospelov:1999mv}. 
It shows for a larger $F_a$, a larger $N$ is needed to achieve good quality. Here we consider two cases $F_a = 10^{9}\,{\rm GeV}$ and $F_a = 10^{12}\,{\rm GeV}$. The former constraint is given by star cooling~\cite{Paul:2018msp}, while in the latter case, the axion is the dominant DM~\cite{Marsh:2015xka}. 
For $F_a = 10^{12}\,{\rm GeV}$, we can derive that the minimum value of $x$ is $7$, which corresponds to
$\delta\bar{\theta}\sim 10^{-26}$. 
Since the number of fermions is small, it is easy to find solutions for fermion charges directly.
And just to be specific, we show one set of many solutions, i.e., $\{-19, -9, -14, -17, 23, -4, 10, -2, 8, -2, 8, -2, 8, -2, 8,$ $1, 5\}$.
Note that the first three are fermion pairs coupled to $\phi_1$.
As for $F_a = 10^{9}\,{\rm GeV}$, the minimum value of $x$ can be $4$, 
which has an extremely high quality, i.e., $\delta\bar{\theta}\sim 10^{-23}$. One set of solutions is $\{-5, -3, -4, -3, 6, 1, 2, 1, 2, 1, 2\}$. Similarly, we have also checked that the $g_3$ will not reach the Landau pole.\\


\noindent \textit{\textbf{Discussion and conclusions.}}---In this Letter, we have proposed a simple framework based on $U(1)_g$ gauge theories with $N$ pairs of chiral fermions. If $N\geq 4$, we have to introduce at least two Higgs bosons $\phi_{1,2}$ to give masses for all fermions, in most of the cases. 
Here we focus on the scenario that two Higgs bosons give all fermion mass through Yukawa interaction.
Therefore, the presence of axions is almost an unavoidable phenomenon in our framework.
Furthermore, we have high-quality axions, including the QCD axion, the Fuzzy DM axion, and the quintessence axion, in large parameter space. 
Their high qualities are guaranteed by the $U(1)_g$  and $\mathbf{Z}_{2N}$ gauge symmetries and therefore
are free from fermion loop and nonperturbative corrections of quantum gravity.

We use this framework to construct quintessence axion and Fuzzy DM through extra leptons, which provide the CS-type interaction that could explain the isotropic cosmic birefringence and have a possible detection channel. 
Finding the fermion $U(1)_g$ charge assignment satisfying Eq.~\eqref{eq:anomaly_cancel} and Eq.~\eqref{eq:higgs_charge} with some fixed $k$ and $l$ is a general mathematical problem, and it is difficult to find the solution directly, especially if the number of fermions is huge. Here we use the trick to shrink the number of free parameters and show some specific solutions for the quintessence axion as well as the Fuzzy DM axion scenarios.
For the QCD axion, we adopt $N=3+2x$ pairs of quarks within this framework as a specific example.
We find that $x=7~(4)$ is the minimum case to provide high-quality QCD axions with $F_a = 10^{12}~(10^{9})\,{\rm GeV}$. Note that we can easily have a QCD axion without the domain wall problem, as long as $2x$ and $3$ are relatively prime numbers.

Also, it's important to note that we are just providing a framework here; it can be extended to many further types of research. 
For example, one could allow higher-order terms to grant fermion mass other than Yukawa interactions. This shall result in a lighter fermion which may show some signature in the thermal history.

One could construct ultra-light bosons with a board mass range under asymmetric charge assignment of fermions in our framework, and their qualities are protected by gauged $U(1)_g$ and $\mathbf{Z}_{2N}$. Such light bosons, $10^{-20}$--$10^{-10}\,{\rm eV}$, may form clouds around astrophysical black holes through superradiance instability~\cite{Brito:2015oca}, which could be further studied by gravitational collider physics~\cite{Baumann:2019ztm}.


We can introduce more than two Higgs bosons and we have many global $U(1)$ symmetries. The spontaneous breaking of these global $U(1)$s generates many axions. Some of them have high quality and some of them do not. In any case, we have multiple axion-like particles. This might be regarded as a generic prediction of our framework.

Another interesting feature of our framework is the appearance of a new massive gauge boson $A'$. After symmetry breaking, the gauge boson mass is $m_{A'} \sim g f$, where $g$ is the $U(1)_g$ gauge coupling and $f=f_1=f_2$. By using the weak gravity conjecture~\cite{Arkani-Hamed:2006emk}, i.e. $g \gtrsim m_{\psi_i}/M_{\rm Pl}$, we can derive $m_{A'} \gtrsim m_{\psi_i}^2/M_{\rm Pl}\sim 100~\text{GeV}$. If consider the mixing between $A'$ and photon, we have $\Gamma_{A'}\sim \alpha \epsilon^2 m_{A'}$ and it can not be the DM unless the mixing parameter $\epsilon$ is extremely small. However, if we identify the $U(1)_g$ with the $B-L$ gauge symmetry, the weak gravity conjecture requires very weak condition, i.e. $g>m_{\nu}/M_{\rm Pl}\simeq 10^{-30}$ with adopting $m_\nu \sim 10^{-3}~\text{eV}$, and the gauge boson can be very light to be the DM. Details of this model were already analyzed by one of the present authors (T.T.Y.) in Ref.~\cite{Lin:2022xbu, Lin:2022mqe}.\\

\begin{acknowledgements}
T. T. Y. is supported in part by the China Grant for Talent Scientific Start-Up Project and by Natural Science Foundation of China (NSFC) under Grant No. 12175134 as well as by World Premier International Research Center Initiative (WPI Initiative), MEXT, Japan.

\end{acknowledgements} 


\bibliography{reference}
\bibliographystyle{apsrev}

\newpage
\onecolumngrid
\fontsize{12pt}{14pt}\selectfont
\setlength{\parindent}{15pt}
\setlength{\parskip}{1em}
\newpage
\begin{center}
	\textbf{\large High-quality axions in a class of chiral $U(1)$ gauge theories} \\ 
	\vspace{0.05in}
	{ \it \large Supplemental Material}\\ 
	\vspace{0.05in}
	{Yu-Cheng Qiu$^1$, Jin-Wei Wang$^1$, Tsutomu T. Yanagida$^{1,2}$}
	\vspace{0.05in}
\end{center}
\centerline{{\it  $^1$Tsung-Dao Lee Institute and School of Physics and Astronomy,}}
\centerline{{\it  Shanghai Jiao Tong University, 520 Shengrong Road, Shanghai, 201210, China}}
\vspace{0.2em}
\centerline{{\it  $^2$School of Physics, University of Electronic Science and Technology of China, Chengdu 611731, China}}
\vspace{0.2em}
\centerline{{\it  $^3$Kavli IPMU (WPI), The University of Tokyo, Kashiwa, Chiba 277-8583, Japan}}
\vspace{0.05in}
\setcounter{page}{1}

In this Supplemental Material, we give a detailed derivation of $[U(1)_a]\times[SU(3)_c^2]$ and $[U(1)_a]\times[SU(2)_L^2]$ anomaly for $N=k+l$ pairs of fermions with ``asymmetric" charge assignment.

\noindent \textit{\textbf{Anomaly for $N=k+l$ with asymmetric charge}}--- The fermion $\psi_i \in (\mathbf{3},\mathbf{1},0)$, with $i =1,2,\cdots,N$, carry $U(1)_g$ gauge charge $\{\alpha_1,~\beta_1,\cdots,~\gamma_1,\cdots,~\delta_1,~\eta_1,\cdots,\sigma_1,\cdots\}$, while for anti-fermion $\overline{\psi}_i \in (\mathbf{3^*},\mathbf{1},0)$ carrys the same $U(1)_g$ gauge charge as $\psi_i$s but not in the same order (see Table. \ref{tab:asymmetriccharges}). As we mentioned in the main text, all fermions' mass terms are generated through the Yukawa couplings, i.e., 
\begin{equation}
    \mathcal{L}_\text{Yukawa} = \sum_{i=1}^k \phi_1 \psi_i \overline{\psi}_i  + \sum_{j = 1}^{l} \phi_2 \psi_j \overline{\psi}_j \;.
\end{equation}
Similarly, we assign the $U(1)_g$ gauge charge of two Higgs $\phi_1$ and $\phi_2$ as
\begin{equation}
-q_1 = \alpha_i+\beta_i = 2\gamma_i,  \quad -q_2 = \delta_i+\eta_i = 2\sigma_i \;.
\end{equation}
Again, here we set $q_1>0$, $q_2<0$. 
By using the $U(1)_g$ gauge anomaly-free condition (see Eq. \eqref{eq:anomali_cancel_1}), we can derive that 
\begin{equation}
    k \frac{q_1}{2}  + l \frac{q_2}{2} = 0 \quad \Rightarrow \quad  -\frac{q_1}{q_2} = \frac{l}{k} = \frac{m}{n}\;.
\end{equation}
Assuming $(q_1,-q_2)=(N_1 l,N_1 k)$, $(l,k)=(N_\text{DW} m, N_\text{DW} n)$, then we have
\begin{equation}
    q_1 = N_1 N_\text{DW} m\;, \quad q_2 = -N_1 N_\text{DW} n\;.
\end{equation}
With the explicit form of $\phi_1$ and $\phi_2$ in the main text we can derive that 
\begin{equation}
    \tilde{a} \rightarrow \tilde{a} + \kappa f_1 q_1\;, \quad \tilde{b} \rightarrow \tilde{b} + \kappa f_2 q_2\;,
\end{equation}
under the $U(1)_g$ transformation, while $\kappa$ is the transformation parameter. 
Knowing that $U(1)_a$ is orthogonal to $U(1)_g$, the transformation of $\tilde{a}$ and $\tilde{b}$ under $U(1)_a$ can be expressed as
\begin{equation}
    \tilde{a} \rightarrow \tilde{a} + \kappa f_2 q_2\;, \quad \tilde{b} \rightarrow \tilde{b} - \kappa f_1 q_1\;,
    \label{eq:u1a}
\end{equation}
which implies that the $U(1)_a$ charge of $\phi_1$ and $\phi_2$ are $f_2 q_2/f_1$ and $-f_1 q_1/f_2$, respectively.
The $[U(1)_a]\times [SU(3)_{\rm c}]^2$ anomaly can be expressed as 
\begin{equation}
    \mathcal{A} = \sum_i^N\left[ U(1)_a^{\psi_i} + U(1)_a^{\overline{\psi}_i}\right] \times \frac{g_s^2}{32 \pi^2} G^{a\mu\nu} \tilde G^a_{\mu\nu}\;,
\end{equation}
where $U(1)_a^{\psi_i}$ and $U(1)_a^{\overline{\psi}_i}$ are $U(1)_a$ charge of $\psi_i$ and $\overline{\psi}_i$, respectively, the $G^{a\mu\nu}$ is the gauge field strength of QCD, and $\tilde G^a_{\mu\nu}$ is its dual. 
The summation of all fermions' $U(1)_a$ charges is 
\begin{equation}
    \sum_{i=1}^{N} \left[ U(1)_a^{\psi_i} + U(1)_a^{\overline{\psi}_i} \right] = - k \frac{f_2 q_2}{f_1} + l \frac{f_1 q_1}{f_2}
    =N_1 N_\text{DW}^2 \frac{\sqrt{f_1^2 m^2+f_2^2 n^2}}{F_a}\;,
\end{equation}
where 
\begin{equation}
F_a = \frac{ f_1 f_2}{\sqrt{f_1^2 m^2 + f_2^2 n^2}}.
\end{equation}
Note that we already used the same notation as in the main text. 
Therefore, $[U(1)_a]\times [SU(3)_{\rm c}]^2$ anomaly is
\begin{equation}
    \mathcal{A} = N_1 N_\text{DW} ^2\frac{\sqrt{f_1^2 m^2+f_2^2 n^2}}{F_a} \frac{g_s^2}{32 \pi^2} G^{a\mu\nu} \tilde G^a_{\mu\nu}\;.
\end{equation}
Besides, performing transformation of  Eq.~\eqref{eq:u1a} and according to Eq.~\eqref{eq:transform}, we can derive that under the $U(1)_a$ transformation,
\begin{equation}
b\to b\;,\quad a \rightarrow a - \kappa N_1 N_\text{DW} \sqrt{f_1^2 m^2+f_2^2 n^2}\;.
\end{equation}
After doing the anomaly matching, the Chern-Simons term should appear in the form of
\begin{equation}
    \mathcal{L} \supset N_\text{DW} \frac{a}{F_a} \frac{g_s^2}{32 \pi^2} G^{a\mu\nu} \tilde G^a_{\mu\nu}\;.
\end{equation}
In particular, when $k$ and $l$ are relatively prime numbers, the domain wall number $N_\text{DW}$ is equal to 1.

Similarly, when $\psi_i \in (\mathbf{1},\mathbf{2},0)$, $\overline{\psi}_i \in (\mathbf{1},\mathbf{2^*},0)$ we can use the same method mentioned above to calculate the $[U(1)_a]\times [SU(2)_{\rm L}]^2$ anomaly, that is
\begin{equation}
  \mathcal{L} \supset N_{\rm DW}  \frac{a}{F_a} \frac{g_2^2} {32\pi^2} W_{\mu\nu}^a\tilde{W}^{\mu\nu a}\;.
    \label{eq:CS_coupling}
\end{equation}

\end{document}